\documentclass[doublecol]{epl2}
\usepackage[T1]{fontenc}
\usepackage[latin9]{inputenc}
\usepackage{amsmath}
\usepackage{esint}

\makeatletter
\@ifundefined{textcolor}{}
{%
 \definecolor{BLACK}{gray}{0}
 \definecolor{WHITE}{gray}{1}
 \definecolor{RED}{rgb}{1,0,0}
 \definecolor{GREEN}{rgb}{0,1,0}
 \definecolor{BLUE}{rgb}{0,0,1}
 \definecolor{CYAN}{cmyk}{1,0,0,0}
 \definecolor{MAGENTA}{cmyk}{0,1,0,0}
 \definecolor{YELLOW}{cmyk}{0,0,1,0}
 }

\title{$^{31}\mbox{P}$ NMR Investigation of the Superconductor LiFeP ($T_c$ = 5 K)}
\shorttitle{$^{31}\mbox{P}$ NMR Investigation of the Superconductor
LiFeP ($T_c$ = 5 K)}

\author{Huiyuan Man, Shengli Guo, Guoxiang Zhi, Xin Gong, Quan Wang, Cui Ding, Yankang Jin and Fanlong Ning\thanks{E-mail: \email{ningfl@.zju.edu.cn}}}
\shortauthor{Huiyuan Man \etal}

\institute{Department of Physics, Zhejiang University - Hangzhou
310027, China }

\pacs{74.70.-b}{Superconducting materials}
\pacs{76.60.-k}{Nuclear magnetic resonance and relaxation}

\abstract{We investigate the static and dynamic spin susceptibility
of the {}``111'' type Fe-based superconductor LiFeP with $T_c$
$\sim$ 5 K through the measurement of Knight shift $^{31}\mbox{K}$
and the spin-lattice relaxation rate $\frac{1}{T_{1}}$ at $^{31}$P
site by nuclear magnetic resonance. The constant $^{31}\mbox{K}$,
small magnitudes of $\frac{1}{T_{1}T}$, along with the resistivity
$\rho$ $\sim$ T$^2$ all point to the weak spin correlations in
LiFeP. $\frac{1}{T_{1}T}$ display small enhancement toward $T_c$,
indicating that the superconductivity is intimately correlated with
the antiferromagnetic spin fluctuations.}

\begin{document}

\maketitle

\section{Introduction}

The discovery of superconductivity with $T_{c}$ = 26 K in the
layered structure $\mbox{La}(\mbox{O}_{1-x}\mbox{F}_{x})\mbox{FeAs}$
($x$ = 0.05 $\sim$ 0.12) \cite{Hosono-LaOFFeAs-2008} has generated
great interest into the research of Fe-based high temperature
superconductors. Over the past six years, hundreds of Fe-based
superconductors have been reported and the list has expanded rapidly
from the original LaFeAsO {}``1111'' structure
\cite{Hosono-LaOFFeAs-2008,Hosono-LaOFeP-2006} to
$\mbox{M}\mbox{Fe}_{2}\mbox{A}\mbox{s}_{2}$ (M stands for alkali
earth metal) {}``122'' family \cite{122-Rotter}, MFeAs (M stands for
alkali metal) {}``111'' family \cite{111-WangXC}, the iron
chalcogenide FeSe {}``11'' family \cite{11-F.C.Hsu}, the {}``42622''
family \cite{42622-H.H.Wen} and the {}``32522'' family
\cite{32522-A.Iyo} etc. Despite the different crystalline
structures, accumulating experimental and theoretical results have
pointed to the unconventional superconducting pair symmetry of these
Fe-based superconductors, and the common physical properties they
shared \cite{Review-G.R.Stewart,Review-H.Hosono}.

LiFeP is one of the prototypical superconductor in the {}``111''
family, which was found to become superconducting below $T_{c} \sim$
6 K \cite{LiFeP-Z.Deng}. In addition to the much lower $T_{c}$,
LiFeP has some properties different from other {}``111''-type
superconductors LiFeAs \cite{LiFeAs-X.C.Wang} and NaFeAs
\cite{NaFeAs-discover}. For example, the measurements of magnetic
penetration depth $\lambda$ have indicated a nodal superconducting
order parameter for LiFeP \cite{Nodal-vs-nodeless}, different from
the fully gapped state observed for LiFeAs \cite{Nodal-vs-nodeless}.
Furthermore, no magnetic or structural transition has been observed
in LiFeP and LiFeAs, but a structural transition at $T_{s}$ = 57 K
\cite{Ts-NaFeAs} and a SDW magnetic transition at $T_{SDW}$ = 45 K
\cite{TSDW-NaFeAs} have been observed in NaFeAs. On the other hand,
it has been shown in LiFeAs by NMR (Nuclear Magnetic Resonance) that
the antiferromagnetic spin fluctuations are strongly enhanced toward
$T_{c}$ \cite{LiFeAs-NMR-W.Q.Yu}. The feature of antiferromagnetic
spin fluctuations enhancement toward $T_{c}$ has been observed in
other Fe-based families
\cite{Ning,BaFe2(AsP)2-NMR-Nakai,FeSe-NMR-T.Imai,LaFeAsOF-NMR-G.Q.Zheng,42622-NMR},
which provides convincing experimental evidences that the
superconductivity is intimately correlated to the antiferromagnetic
spin fluctuations. To the best of our knowledge, no NMR
investigations of LiFeP have been reported. It will be interesting
to investigate the spin dynamics and examine if such feature exists
in the superconducting LiFeP.

In this paper, we conduct NMR measurement on a polycrystalline
sample in the paramagnetic state. We measured the static
susceptibility and the spin dynamics through the Knight shift and
$\frac{1}{T_{1}T}$ measurements at $^{31}$P site, respectively. We
found that the static susceptibility is temperature independent
throughout the measured temperature range of 4.2 K and 280 K. While
the magnitude of antiferromagnetic fluctuations is almost an order
of magnitude smaller than those of LiFeAs \cite{LiFeAs-NMR-W.Q.Yu},
a weak enhancement toward $T_{c}$ is observed. This indicates that
the superconductivity is intimately correlated with the
antiferromagnetic spin fluctuations despite the lower $T_{c}$ of
LiFeP.

\section{Experiments}

The LiFeP polycrystalline specimens were synthesized by the solid
state reaction method. The pallets of mixed high-purity Fe (99.9 \%)
and P (99 \%) powders (Alfa Aesar) were sealed in an evacuated
quartz tube and heated to 800 $^{\circ}$C for 10 hours to prepare
the intermediate product FeP. FeP were then mixed with Li ingots
(Alfa Aesar, 99.9\%) with nominal concentration and heated to 800
$^{\circ}$C for 30 h. The specimens were then cooled down to room
temperature with the furnace shutting off. The handling of materials
were performed in a high-purity argon filled glove box (the
percentage of O$_{2}$ and H$_{2}$O $\leq$ 0.1 ppm), to protect it
from exposing to air. The color of the sample was shinny black,
indicating the good crystallization.

The polycrystals were characterized by the X-ray powder diffraction
and the dc magnetization with a Quantum Design superconducting
quantum interference device (SQUID). The temperature dependence of
electrical resistivity was measured on a thin bar-shaped sample
(3.78 mm$\times$1.02 mm $\times$1.08 mm) in a Cryogenic Mini-CFM
system by a standard four-probe method. We conducted the
$^{31}\mbox{P}$ NMR measurements by using the standard pulsed NMR
techniques. We obtained the NMR spin echo signal by applying
90$^o$-180$^o$ pulses in a fixed external field of $B_{ext}$ = 4.65
Tesla.

\section{Results and discussion}

\begin{figure}[!htpb] \centering
\centering
\includegraphics[width=3in]{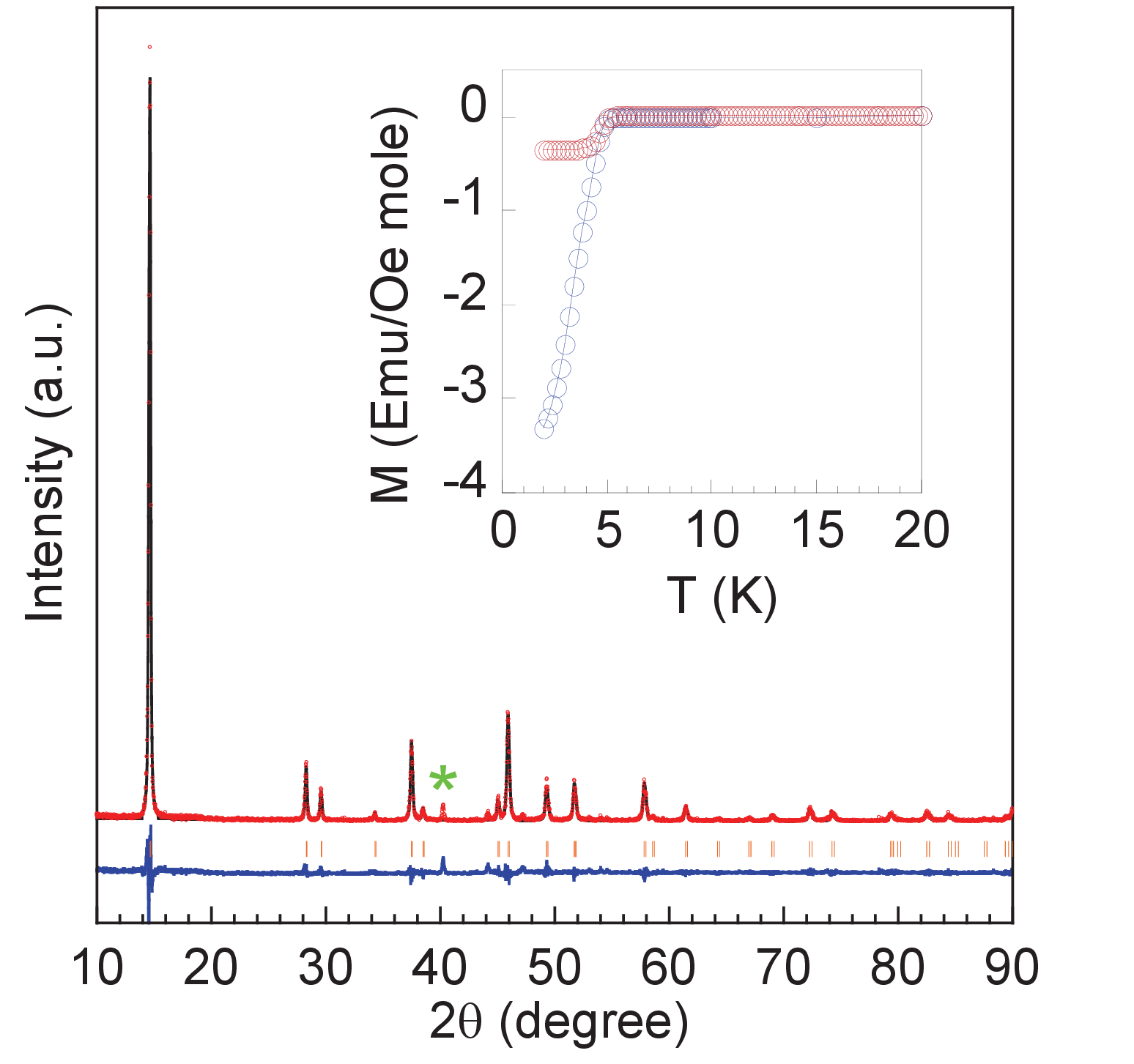}
\caption{(Color online) Powder X-ray diffraction pattern of LiFeP.
Star marks the small amount of $\mbox{Fe}_{2}\mbox{P}$ impurity. The
inset displays the dc susceptibility of LiFeP sample in both ZFC and
FC mode, and the diamagnetic signals indicate that $T_c$ $\sim$ 5
K.} \label{Fig.1}
\end{figure}

In Fig. 1, we show the powder X-ray diffraction pattern of LiFeP
polycrystal. The diffraction peaks can be well indexed into a
$\mbox{Cu}_{2}\mbox{Sb}$ type tetragonal structure with $P4/nmm$
symmetry \cite{LiFeP-Z.Deng}, which is the same as the other two
\textquotedblleft{}111\textquotedblright{} type iron-based
superconductor LiFeAs \cite{LiFeAs-X.C.Wang} and NaFeAs
\cite{NaFeAs-discover}. A careful inspection indicates that small
amount of $\mbox{Fe}_{2}\mbox{P}$ exists at
$2\theta=40.28$$^{\circ}$. $\mbox{Fe}_{2}\mbox{P}$ is a
ferromagnetic material with Curie temperature more than 200 K
\cite{magnetic-Fe2P}. This small amount of impurity will not affect
our NMR measurements of the intrinsic properties of LiFeP since NMR
is a local and site-selective microscopic probe and we conducted NMR
measurements at $^{31}$P site of LiFeP. The lattice constants are
$a=3.6938${\AA} and $c=6.0446${\AA}, which are consistent with the
previous reported values \cite{LiFeP-Z.Deng}. Compared to the values
of $a=3.9494${\AA} and $c=7.0396${\AA} for NaFeAs
\cite{NaFeAs-discover}, and $a=3.77${\AA} and $c=6.36${\AA} for
LiFeAs \cite{LiFeAs-X.C.Wang}, both the ab-plane and the c axis of
LiFeP shrink to some extent. The smaller lattice constants of LiFeP
have been attributed to much smaller atomic size of Li and P atoms
than that of Na and As. In the inset of Fig. 1, we show the dc
magnetic susceptibility of LiFeP sample measured in both ZFC and FC
condition with an applied field of 20 Oe. The diamagnetic signals
confirm the superconductivity takes place at $\sim$ 5 K. The
superconducting volume fraction reaches 100\% for ZFC mode at 2 K
after the correction of the demagnetizing factors. The bulk
superconductivity has also been confirmed through the observation of
abrupt change of conduction frequency in a NMR coil.

In Fig. 2, we show the temperature dependence of electrical
resistivity of LiFeP. The resistivity data indicates a sharp
superconducting transition temperature at $\sim$ 5 K. The residual
resistivity in our specimen is 5.7 $\mu\Omega$.cm, which is smaller
than the value of previous reports
\cite{LiFeP-Z.Deng,Tc-pressure-LiFeP}, indicating the good quality
of our polycrystalline specimen. We plot the resistivity versus
$T^{2}$ in the temperature range of 5 K and 40 K, and a linear
dependence has been observed. We do not observe anomalies
corresponding to the SDW transition or structural transition that
have been observed in the NaFeAs superconductor
\cite{NaFeAs-discover,Ts-NaFeAs,TSDW-NaFeAs}. This situation is
similar to the case of LiFeAs \cite{LiFeAs-X.C.Wang} where no
anomalies have been observed in the curves of both electrical
resistivity and magnetic susceptibility.

\begin{figure}[!htpb] \centering
\includegraphics[width=3in]{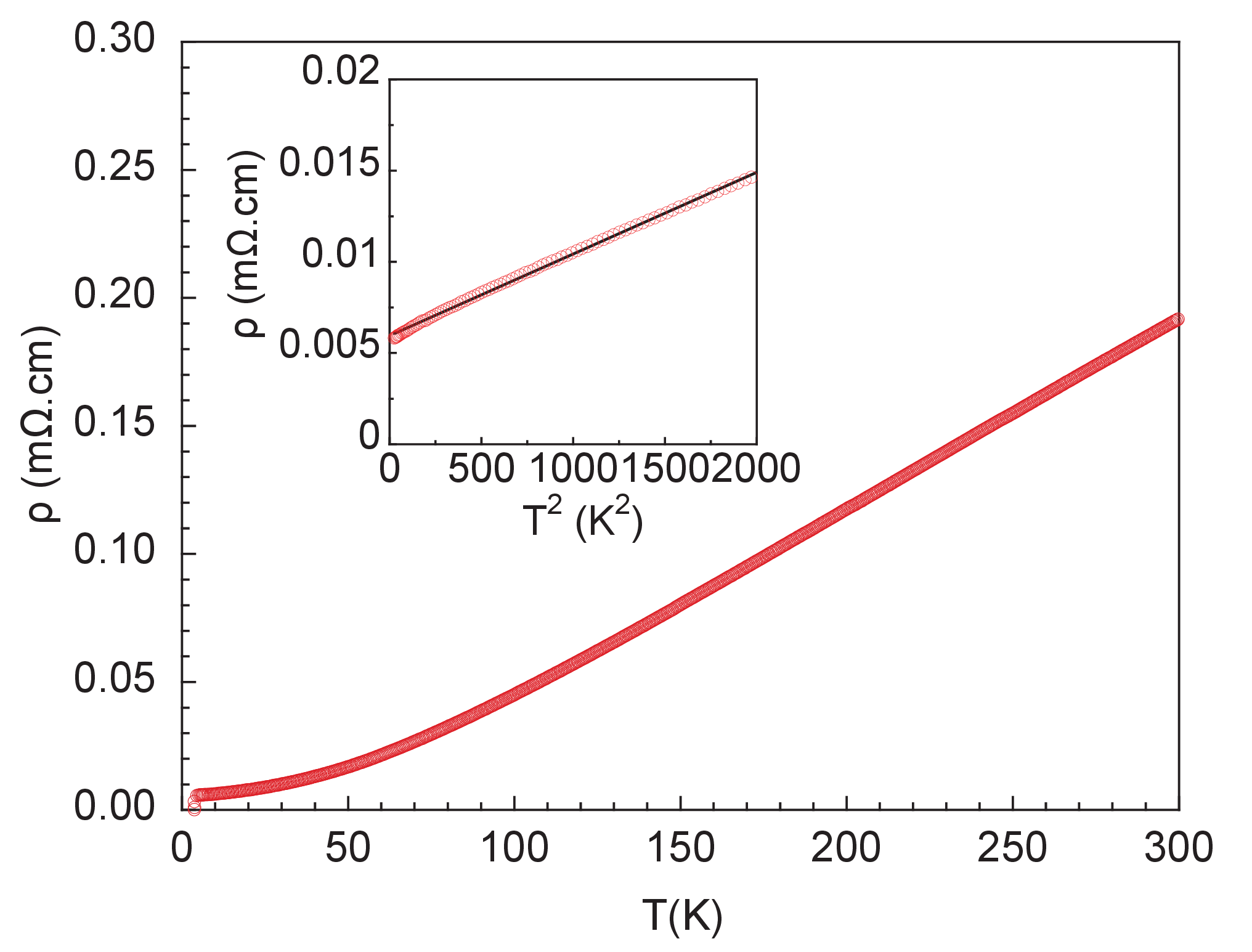}
\caption{(Color online) Temperature-dependent electrical resistivity
for LiFeP polycrystalline sample, showing superconducting transition
at $\sim$ 5 K. The inset shows the expanded $\rho\left(T^{2}\right)$
data, suggesting a linear behavior at low temperature; The black
solid line is a fitting line with the function
$\rho=\rho_{0}+AT^{2}$ with $\rho_{0}$ = 0.0057 $m\Omega$.cm and $A
= 4.5 \times 10^{-6}m\Omega.cm/K^2$.} \label{Fig.2}
\end{figure}

\begin{figure}[!htpb] \centering
\includegraphics[width=3in]{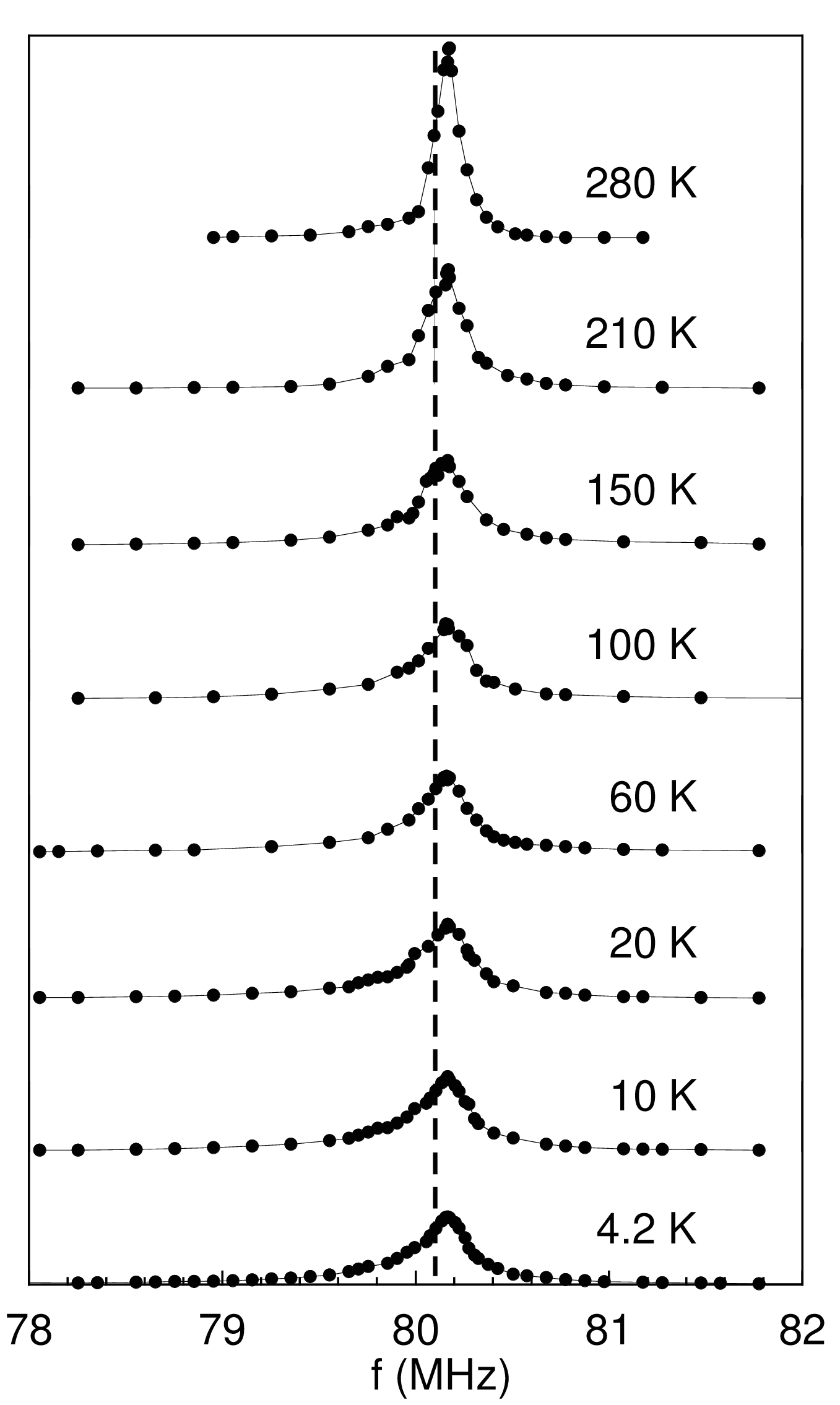}
\caption{$^{31}\mbox{P}$ lineshapes of LiFeP measured over a wide
temperature range from 4.2 K to 280 K under a field of 4.65 Tesla.
The dashed line marks the position of $^{31}\mbox{K}$ = 0.}
\label{Fig.3}
\end{figure}

We carried out the $^{31}\mbox{P}$ NMR lineshapes measurements under
the field of $B_{ext}$ = 4.65 Tesla. This field is much higher than
the second critical field $\sim$ 2 Tesla of LiFeP
\cite{Mossbauer-LiFeP,contrasts-LiFeP-LiFeAs}, and therefore it is
in the paramagnetic state. $^{31}\mbox{P}$ has a nuclear spin
$I=\frac{1}{2}$ with a gyromagnetic ratio of
$^{31}\gamma_{n}/2\pi=17.235$ MHz/Tesla. We would observe a single
resonance frequency for each individual $^{31}\mbox{P}$ site since
there is no nuclear quadrupole interaction. In Fig. 3, we display
the temperature dependence of $^{31}\mbox{P}$ lineshapes for LiFeP
from 280 K to the base temperature of 4.2 K. Only one resonance
frequency is observed at $f \sim 80.166$ MHz, indicating that all
$^{31}\mbox{P}$ atoms are in the same magnetic and electrical
environment. For a naked nuclei, the center of resonance frequency
is $f_{0}=({}^{31}\gamma_{n}/2\pi)\times B_{ext}=80.143$ MHz. It is
the local field arising from the hyperfine interaction between
$^{31}\mbox{P}$ nuclei and surrounding electrons that shift the
resonance frequency $f_{0}=80.143$ MHz to the slightly larger value,
$f\sim 80.166$ MHz. We obtain the Knight shift $^{31}\mbox{K}$ for
each temperature by using the formula $^{31}\mbox{K} = (f-f_0)/f_0
\times 100\%$, where $f$ is the peak frequency of the lineshape. The
temperature dependence of $^{31}\mbox{K}$ is plotted in Fig. 5(a).
$^{31}\mbox{K}$ of LiFeP is only $\sim$ 0.03 \%, which is much
smaller than the $^{75}$K $\sim$ 0.18 \% at $^{75}$As of LiFeAs
\cite{As-NMR-NQR-LiFeAs-German,As-NQR-NMR-LiFeAs-Korringa-ZhengGQ}
and NaFeAs \cite{Na-As-NMR-NaFeAs-JPSJ}. More interestingly,
$^{31}\mbox{K}$ hardly changes in the measured temperature range.
This situation is different from the case of $^{75}\mbox{K}$ of
LiFeAs, where $^{75}\mbox{K}$ smoothly decreases with the decreasing
temperature
\cite{As-NMR-NQR-LiFeAs-German,As-NQR-NMR-LiFeAs-Korringa-ZhengGQ}.

\begin{figure}[!htpb] \centering
\includegraphics[width=3in]{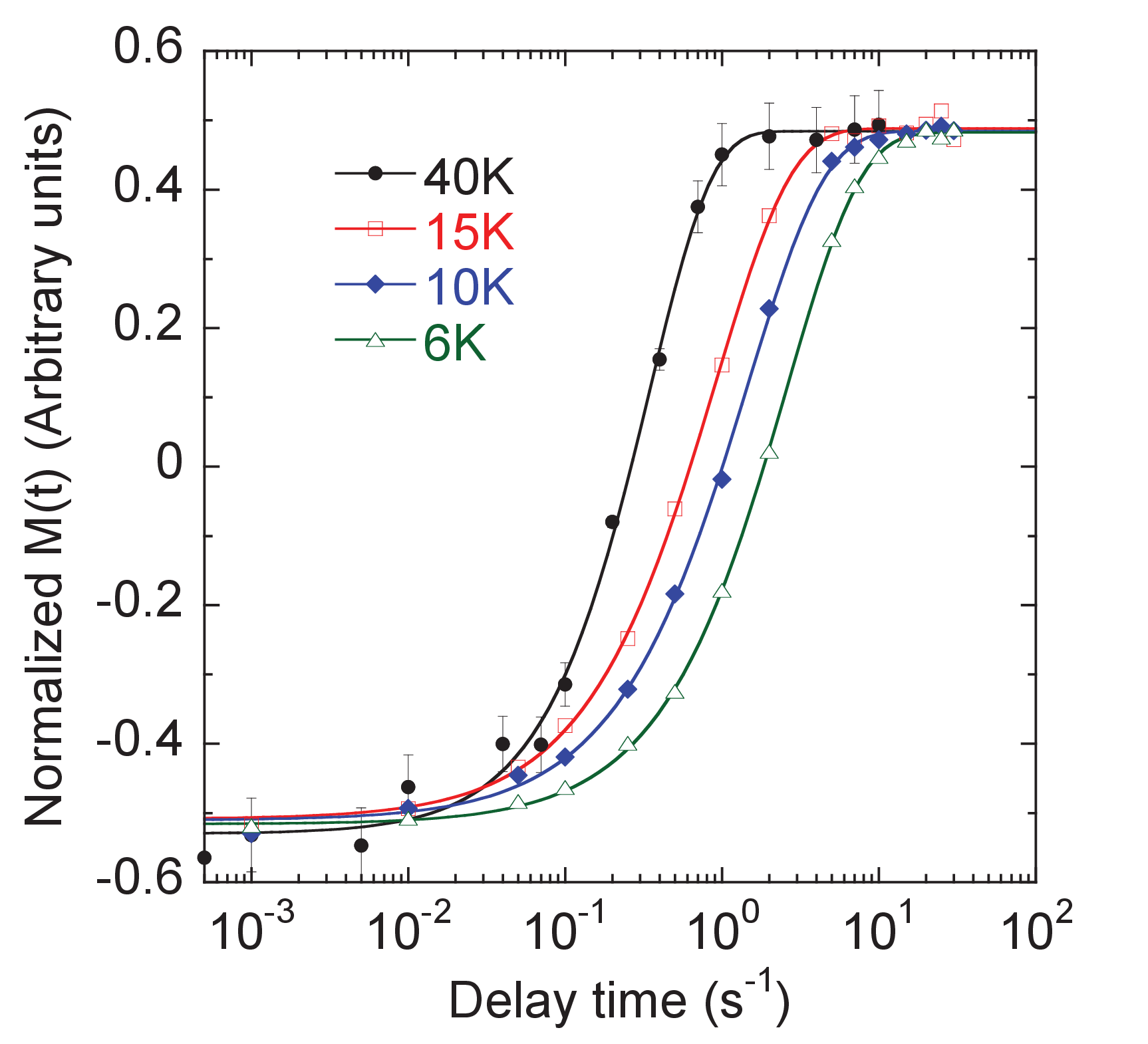}
\caption{(Color online) Typical nuclear spin recovery curves M(t)
after an inversion pulse for $^{31}$P. Solid curves represent
appropriate fits to determine $T_1$ with the stretched exponential
time dependence.} \label{Fig.4}
\end{figure}

To obtain the insight of the spin dynamics, we measure the
spin-lattice relaxation rate $\frac{1}{T_{1}}$ at the peak frequency
of $^{31}\mbox{P}$ NMR lines. $T_{1}$ represents the time scale
during which the nuclear spins return to its thermal equilibrium
after the absorption of the inverted radio-frequency pulse. The
recovery of the nuclear magnetization after the inversion pulse,
M(t), was fitted to a stretched exponential equation, $M(t) = M_0[1-
A exp[-(t/T_1)^{\alpha}]$ where $M_0, A, T_1$ and $\alpha$ are the
free parameters. We show typical nuclear spin recovery curves of the
LiFeP sample for $^{31}$P site in Fig. 4. The exponent $\alpha$
varies smoothly from $\sim$ 1.0 at 280 K to $\sim$ 0.9 at 4.2 K.
Theoretically, the spin contribution to $\frac{1}{T_{1}}$ may be
written using the imaginary part of the dynamical electron spin
susceptibility ${\chi''({\bf q},f)}$ as \cite{Moriya},
 \begin{equation}
          \frac{1}{T_{1}} =
          \frac{2\gamma_{n}^{2}k_{B}T}{g^{2}\mu_{B}^{2}}\sum_{{\bf q}} | A({\bf q}) |^{2} \frac{\chi''({\bf
    q},f)}{f},
\label{T1}
\end{equation}where $A({\bf q})$ is the hyperfine form factor \cite{Moriya} and $f$ the measured frequency.
$T_1$ can be as long as $\sim$ 250 seconds for non-magnetic
insulators such as the direct gap semiconductor LiZnP, and become
five orders shorter to $\sim$ 2.5 milliseconds when strong magnetic
fluctuations exist \cite{Ding}.

\begin{figure}[!htpb] \centering
\includegraphics[width=3in]{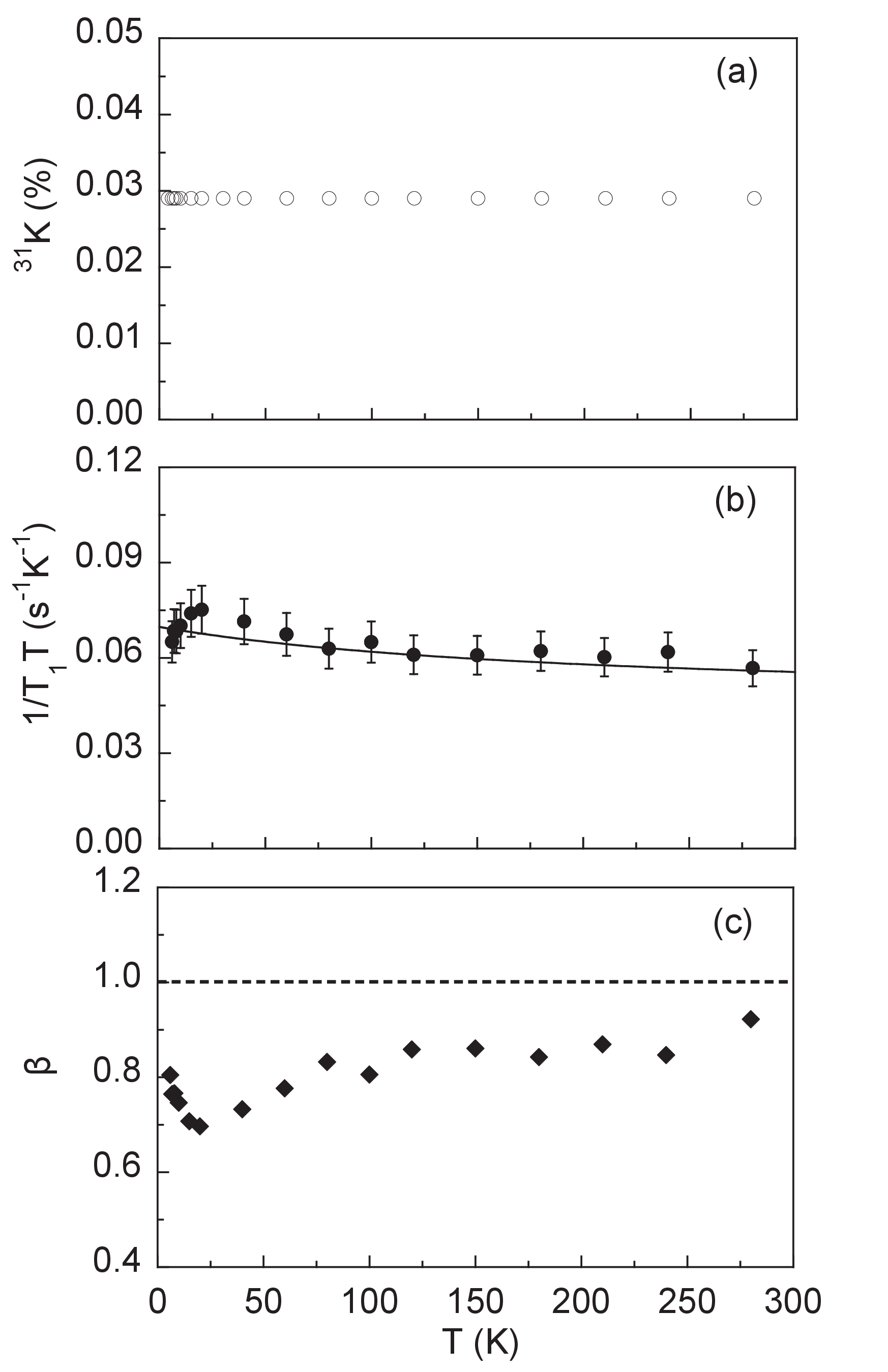}
\caption{(a) The temperature dependence of $^{31}\mbox{P}$ NMR
Knight shifts. (b) $T$-dependent $\frac{1}{T_{1}T}$ for LiFeP. The
solid line is a guide for eyes. (c) The plot of Korringa factor
$\beta$ versus temperature; Horizontal dotted line indicates an
expected value $\beta = 1$ for noninteracting electrons.}
\label{Fig.5}
\end{figure}

We show the $T$-dependence of the spin-lattice relaxation rate
$\frac{1}{T_{1}}$ divided by T, $\frac{1}{T_{1}T}$ in Fig. 5(b). We
note that the results of $\frac{1}{T_{1}T}$ hardly change even if we
fix the exponent $\alpha$ equal to 1.0 throughout the entire
temperature range. The value of $\frac{1}{T_{1}T}$ at
$^{31}\mbox{P}$ site of LiFeP is $\sim$ 0.065 s$^{-1}$K$^{-1}$,
which is almost an order smaller than $\sim$ 0.45 s$^{-1}$K$^{-1}$
at $^{75}\mbox{As}$ sites of LiFeAs
\cite{LiFeAs-NMR-W.Q.Yu,As-NMR-NQR-LiFeAs-German,As-NQR-NMR-LiFeAs-Korringa-ZhengGQ,
As-NMR-LiFeAs-Korringa-Slovenia}, implying much weaker spin
fluctuations in the LiFeP system. This result also indicates that
the electron correlation is much weaker in LiFeP than LiFeAs, and is
consistent with significant mass enhancement in LiFeAs than LiFeP
observed in high field quantum oscillations \cite{Putzke-PRL}. In
addition, we find that $\frac{1}{T_{1}T}$ display a slight
enhancement toward the low temperature. This feature is
qualitatively similar to the case of LiFeAs
\cite{LiFeAs-NMR-W.Q.Yu}, indicating that antiferromagnetic spin
fluctuations do exist in LiFeP although they are weaker.

We can also use the Korringa relation,
$T_{1}TK_{s}^{2}=\frac{\hbar}{4\pi
k_{B}}\frac{\gamma_{e}^{2}}{\gamma_{n}^{2}}\beta$, to evaluate
quantitatively the strength of the electron correlations. Where
$\gamma_{e}$ and $\gamma_{n}$ are the electron and nuclear
gyromagnetic ratios, respectively. $K_{s}$ is the spin
susceptibility extracted from the Knight shift data in Fig. 5(a).
The Korringa factor $\beta$ reflects the magnitude of spin
correlations \cite{Korringa-beta}. Usually $\beta$ is equal to 1 for
a noninteracting system, and strong ferromagnetic correlations give
$\beta\gg1$, while strong antiferromagnetic correlations give
$\beta\ll1$. We show the calculated $\beta$ in Fig. 5(c). The values
of $\beta$ in the whole measured temperature range are smaller than
1, indicating the existence of the antiferromagnetic fluctuations in
LiFeP. $\beta$ is close to 1, which is much larger than $\beta$
$\sim$ 0.2 $\sim$ 0.5 in LiFeAs
\cite{As-NQR-NMR-LiFeAs-Korringa-ZhengGQ,
As-NMR-LiFeAs-Korringa-Slovenia}, again indicating that the weaker
spin correlations in LiFeP.

A close inspection of the $\frac{1}{T_{1}T}$ curve indicates that a
small hump appears at $T$ $\sim$ 20 K, suggesting some magnetic
instabilities exist around this temperature. Currently we do not
know the origin of this magnetic instability since no anomalies have
been detected around this temperature in both dc magnetic
susceptibilities and the electrical resistivity. A brave assumption
is that this hump is related to the SDW magnetic transition which
has been absent in LiFeP and LiFeAs. A possible scenario is that Li
concentrations affects the ground state of
$\mbox{Li}_{1\pm\delta}\mbox{FeP}$, and a precise controlling of Li
concentration may unveil this mystery.

\section{Summary and Conclusion}

In summary, we investigated the static and dynamic spin
susceptibility of the superconducting LiFeP in the paramagnetic
state by NMR measurement at $^{31}$P site. Through the measurement
of electrical resistivity, Knight shift and $\frac{1}{T_{1}T}$, we
found that $\rho$ $\sim$ T$^2$, $^{31}\mbox{K}$ $\sim$ constant and
Korringa ratio $\beta$ $\sim$ 0.8, which all point to the weak spin
correlations in LiFeP. Our results indicate that the
antiferromagnetic spin fluctuations are slightly enhanced toward
$T_c$ although their magnitudes are much weaker than those of
LiFeAs. This on one hand indicates that antiferromagnetic
fluctuations are important for the superconductivity in LiFeP, and
on the other hand may explain the much lower $T_c$ of LiFeP than
that of LiFeAs. We also detected an magnetic instability around
$\sim$ 20 K, which may be related to the different ground states
arising from Li off-stoichiometry. We are applying electrochemical
method to precisely control the amount of Li concentration in LiFeP
and LiFeAs to elucidate their mysterious ground states.

\acknowledgments

The work at was supported by National Basic Research Program of
China (No.2011CBA00103, 2014CB921203), NSF of China (No.11274268).

\end{document}